\documentstyle[11pt,newpasp,twoside,epsf]{article}
\def\edcomment#1{\iffalse\marginpar{\raggedright\sl#1\/}\else\relax\fi}
\reversemarginpar

\begin{document}
\title{Evolutionary Memory in Binary Systems?}
\author{Netzach Farbiash \& Raphael Steinitz}
\affil{Ben-Gurion University of the Negev, P.O. Box 653 Beer-Sheva
84105 Israel}

\begin{abstract}
Correlation between the spins (rotational velocities) in binaries
has previously been established. We now continue and show that the
degree of spin correlation is independent of the components'
separation. Such a result might be related for example to Zhang's
non-linear model for the formation of binary stars from a nebula.
\end{abstract}

\section{Introduction}
Synchronization is common among members of close binary systems.
Theoretical arguments for such synchronization have been advanced
by Zahn (1970, 1975, and 1977). Empirical results by Levato
(1976), Giuricin G. et al. (1984; 1984), and others, seem to agree
with Zahn's model.

As a binary system evolves, the original angular momentum of the
accreting mass is shared between orbital and spin angular
momentum, provided no external perturbations is present. Thus, we
expect that spin angular momenta of the components are roughly
parallel to the orbital one. In that case, the measured $v \sin{i}
$ values should be correlated.

Indeed, Steinitz $ \&$ Pyper (1970) found that there is such a
correlation between the projected spins of the members in binary
systems. Subsequent work by Farbiash $ \&$ Steinitz (2003) yielded
similar results for an extended sample (1010 binary systems). Data
selection and restrictions imposed on choice of the binary systems
is given there. The basic conclusion arrived at was that

\begin{equation}
v_{1}\ sini_{1} \cong v_{2} \ sini_{2}, \
\end{equation}

which can be understood either as:

\begin{enumerate}
    \item $v_{1} \ll\ v_{2}$ while $sini_{1} \gg\
sini_{2}$,

or:
    \item $v_{1} \simeq\ v_{2}$ as well as $sini_{1}
\simeq\ sini_{2}$.
\end{enumerate}

Since $i_1$ and $i_2$ are angles depending on the observer, the
probability of the first relation is extremely small. We accept
the second relation, interpreting it as twofold meaning: Spin axes
of members in binary systems are roughly parallel, and also
rotational speeds are correlated.

Tidal interaction in close binary systems is an important process,
therefore we expect this result. Remembering, however, that this
interaction is strongly dependent on distance between the
interacting stars (Zahn, 1977), we ask whether tidal interaction
is the only process bringing about spin correlation. If it is
indeed the only process causing spin correlation, then this
correlation will diminish with increasing separation distance.
This is the subject of the current investigation.

\section{Data}
Choice of the data to be examined has been described originally by
 Farbiash \& Steinitz (2003). For brevity we mention the salient
points:

\begin{enumerate}
    \item Spectral type of both components is earlier
than F0 (slow rotation of stars later than F0 would automatically
simulate correlation).
    \item Giants and Supergiants are excluded since they may have lost their original
rotational velocities.
    \item Multiple systems including more than two stars are also
excluded.
\end{enumerate}

To ensure that the looked after correlation is not accidental due
to proximity in spectral type, we have previously defined two
extra sets of artificial binaries AB(Artificial Binaries), and
ABR(Artificial Binaries, Restricted). These were obtained by
shuffling the original components (regarded as single stars), and
further restricting to a very narrow range in spectral type.

In addition to these samples, we now define a new subset,
VB(Visual Binaries) containing only visual binaries whose
separation between the components is known (Hartkopf \& Mason,
2003). This set contains 33 systems.

\section{Results and Conclusions}

In fig.1 we plot the projected rotational velocity of one
component versus the other component for all samples: AB, ABR, RB,
and VB. The relevant part of the figure is the plot for the VB
sample. Also, we plot in fig.2 the projected
 rotational velocity differences against the components' separation.

From fig.1 we learn that:

\begin{itemize}
    \item Spin correlation in visual binaries is indeed present.
\end{itemize}

And from fig.2 we see that:

\begin{itemize}
    \item Spin correlation \textbf{does not depend on separation
    distance}.
\end{itemize}

\begin{figure}
\plotone{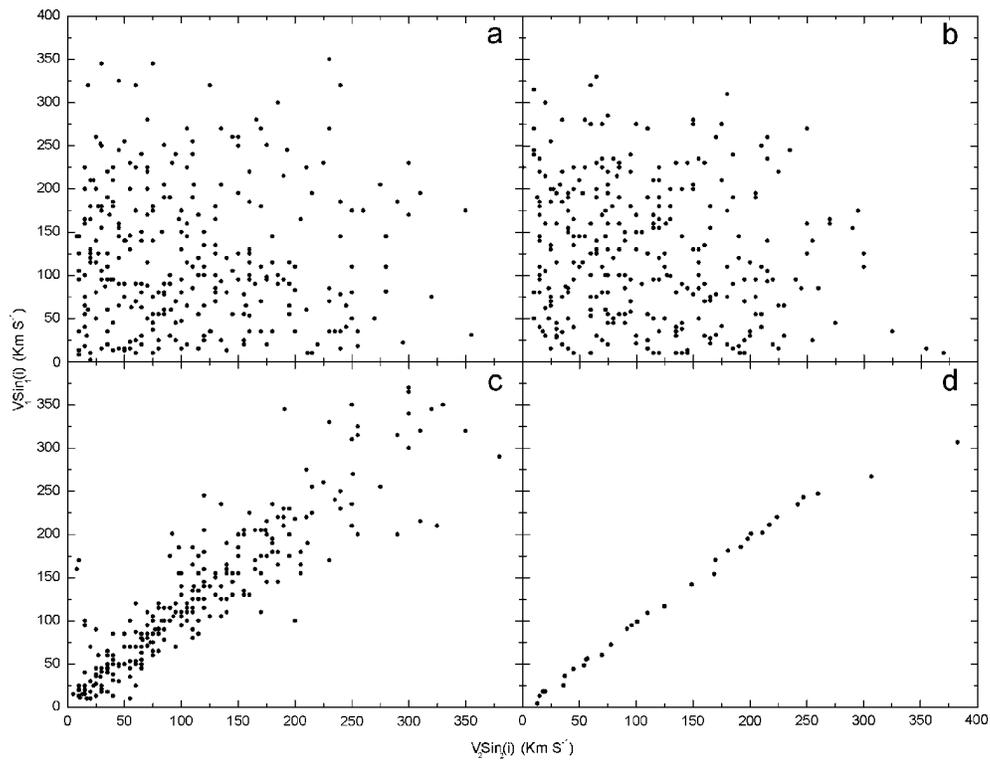}
%\plotfiddle{farbiash_fig1.eps}{4.037in}{0}{100}{100}{0}{0}
 \caption{Projected rotational
velocity of one component vs. the other component for samples AB
(a), ABR (b), RB (c), and VB (d). (Fig 1a,1b, and 1c appeared in
Farbiash \& Steinitz, 2003)}
\end{figure}

\begin{figure}
\plotone{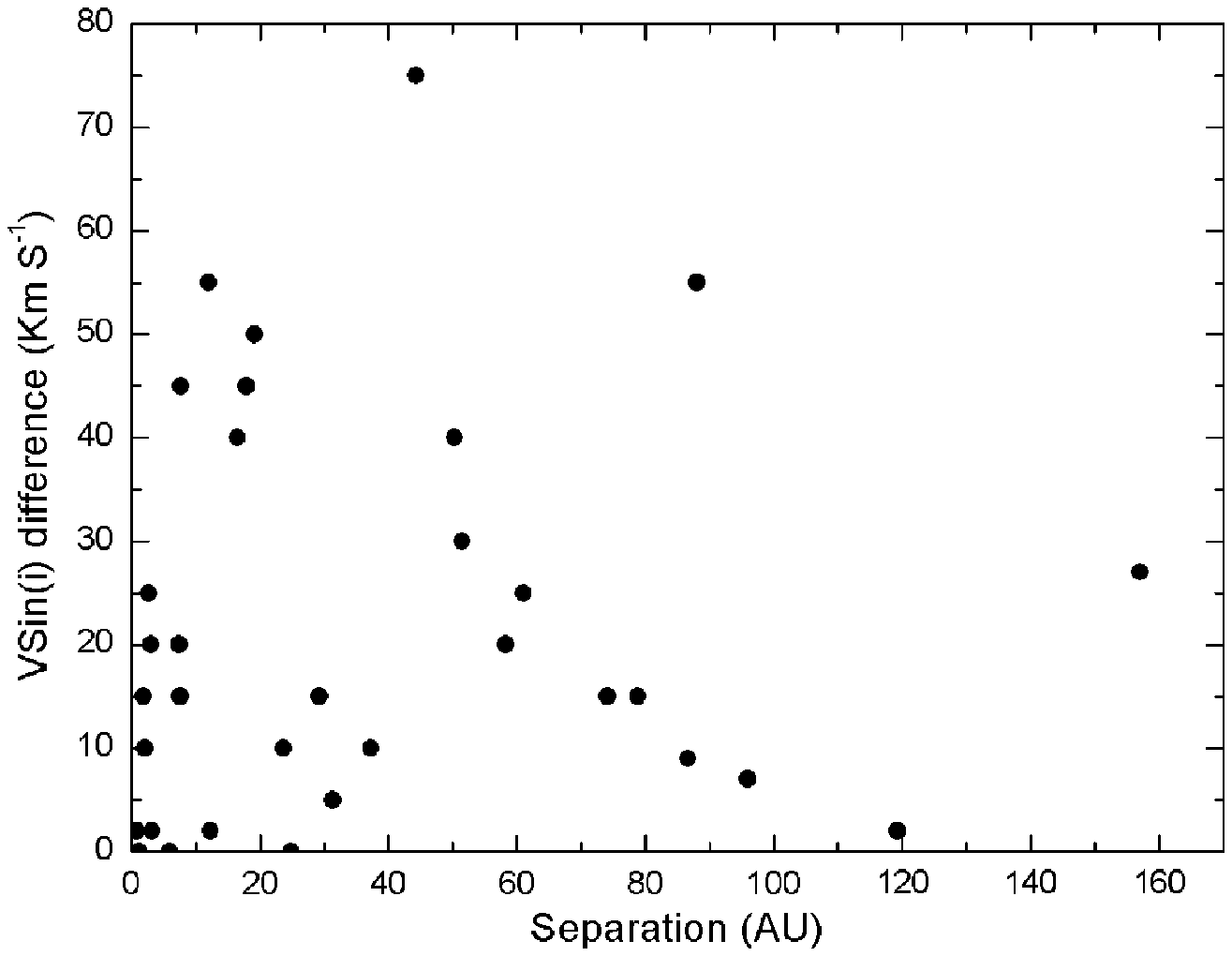}
%\plotfiddle{farbiash_fig2.eps}{3in}{0}{100}{100}{0}{0}
\caption{Projected
 rotational velocity differences vs. the separation of components for sample VB(Visual Binaries).}
\end{figure}

"Memory" of initial conditions in close binary system are
obviously erased due to tidal interaction. As this interaction
becomes negligible for visual binaries, spin correlation just
demonstrated should therefore be due to other mechanisms, which
possibly do reflect initial conditions. For example, Zhang (2000)
gave a non-linear model for the evolution of binary stars from a
nebula, in which spin correlation is retained. For the current
investigation the sample available to us is small - 33 visual
binaries. A firmer base, an enlarged sample of visual binaries,
would enable us to obtain more significant results pertaining to
spin correlation, and thus obtain a better idea of the "memory" of
initial conditions. We encourage therefore observers to collect
$v\sin i$ values of a large sample of visual binary systems.

\end{document}